\documentclass[twocolumn,nofootinbib,prd,aps,
tightenlines]{revtex4}
\usepackage{color}
\usepackage{bm}
\usepackage{epsfig}
\usepackage{pstricks}
\def\be{\begin{eqnarray}}\def\ba{\begin{eqnarray}}
\def\bi{\bibitem}
\def\ee{\end{eqnarray}}\def\ea{\end{eqnarray}}
\def\del{\partial}
\def\ben{\begin{enumerate}}\def\een{\end{enumerate}}
\def\bitem{\begin{itemize}}\def\eitem{\end{itemize}}
\def\lsim{\mathrel{\rlap{\lower3pt\hbox{\hskip1pt$\sim$}}
     \raise1pt\hbox{$<$}}} 
\def\gsim{\mathrel{\rlap{\lower3pt\hbox{\hskip1pt$\sim$}}
     \raise1pt\hbox{$>$}}} 
\def\la{\langle}\def\ra{\rangle}
\def\calL{\cal L}\def\calO{\cal O}

\def\pr{Phys. Rev.}\def\prl{Phys. Rev. Lett.}
\def\np{Nucl. Phys.}\def\pl{Phys. Lett.}

\def\thetap{\Theta^+}
\begin{document}
\title{Pentaquarks, Skyrmions and the Vector Manifestation of Chiral
Symmetry\footnote{Based on talks given at {\it the Hanyang-KIAS
Workshop on Multifaceted Skyrmions and Effective Field Theory},\
KIAS, Seoul, Korea, 25-27 October 2004 and at {\it the
International Workshop on Dynamical Symmetry Breaking}, Nagoya
University, Nagoya, Japan, 21-22 December 2004.}}
\author{Mannque Rho$^a$}\email{rho@spht.saclay.cea.fr}
\affiliation{$^a$ Service de Physique Th\'eorique, CEA Saclay,
91191 Gif-sur-Yvette, France}
\date{\today}
\begin{abstract}
The structure of the pentaquark baryon $\Theta^+$ is discussed in
terms of a $K^+$-skyrmion binding where the skyrmion arises as a
soliton in hidden local symmetry approach to low-energy hadronic
physics which may be considered as a holographic dual to QCD. The
``vector manifestation" of chiral symmetry encoded in the
effective theory Wilsonian-matched to QCD is proposed to play an
important role in the binding. Among the options available for
understanding the pentaquark structure is the intriguing
possibility that the $\thetap$ is a Feshbach resonance generated
by the solitonic matter that drives the Wess-Zumino term that in
the presence of $K^\star$ acts like a magnetic field.

\end{abstract}
\maketitle

\renewcommand{\thefootnote}{\arabic{footnote}}
\setcounter{footnote}{0}
\section{Pentaquarks}
The prediction~\footnote{The Skyrme model prediction for this
state has a long history starting with the earliest one by
Praszalowicz in 1987~\cite{prasz}.} in 1997 by Diakonov, Petrov
and Polyakov (DPP)~\cite{DPP} of the $S=+1$ baryon denoted
$\thetap$ at 1540 MeV with a narrow width $\lsim 20$ MeV and the
subsequent experimental confirmation of the state published in
2003 by Nakano et al~\cite{Nakano} with mass $1540\pm 10$ MeV and
a width $<25$ MeV have spurred a huge activity both in experiment
and in theory. What was remarkable in the DPP work was that it was
an elegantly simple prediction based on a clearly defined
theoretical framework of chiral solitons and that considering the
qualitative nature of the soliton model, the prediction was
confirmed surprisingly quantitatively. Since then there have been
a flurry of activities with conflicting results~\cite{review},
some pro and some con, both in theory and in experiment and the
verdict on the existence or non-existence of the $\thetap$ is yet
to come out from the experimental side. Fortunately a clear-cut
experimental confirmation or refutation is expected to be
forthcoming in the near future.

The discovery of pentaquark baryons will undoubtedly be a boon for
hadron spectroscopy. I would like to further suggest that
independently of whether or not the $\Theta^+$ in question exists
in nature, the development brings out a highly intriguing
theoretical issue tied to chiral symmetry and its spontaneous
breaking in matter. This is the topic of my talk here.

For the discussions that follow, I will assume that the $\thetap$
exists. Even if it is proven to be non-existent by forthcoming
experiments, what I will discuss will still be relevant for other
issues I will make reference to, so in either case, my discussions
will not be futile. For definiteness I will focus solely on the
singlet member of the $\overline{10}$ multiplet to which the
$\thetap$ is assumed to belong. For the physically more meanigful
spectroscopy of pentaquark baryons, a lot more work will be
needed. Also I shall not touch on quark-model descriptions,
focusing instead on the ``dual" skyrmion description.
\section{Skyrmions}
When the prediction for the $\thetap$ was first made in 1980's and
later in 1990's, it was considered to be a unique prediction of
the chiral soliton model not shared by quark models in that it
naturally arose from the model on the same footing as the baryons
described in terms of $N_c=3$ quarks in the quark model. In the
naive quark picture, the $\thetap$ is ``exotic" since it requires
at the minimum four quarks and one antiquark. But of course there
is nothing sacred in this naive picture. Why should there not be
baryons that carry more than $N_c$ valence quarks just as in
nuclear physics more than valence particles are involved in
nuclear spectroscopy? So in this sense, there is nothing really
``exotic" about the five-quark structure of a baryon in QCD. In
fact, in the large $N_c$ limit, $all$ baryons in the skyrmion
model can be matched one-to-one to constituent quark
model~\cite{manoharN}. However for convenience of understanding
what we mean, I will refer to the standard $N_c$ quark baryons as
``non-exotic" and to all others ``exotic."
\subsection{The skyrmion Lagrangian}
The skyrmion prediction typified by DPP was based on the usual
Skyrme Lagrangian
 \be
{\calL}&=&\frac{f^2}{4}{\mbox Tr}\del_\mu U\del^\mu U^\dagger
+\frac{1}{32 e^2}{\mbox Tr}[U^\dagger\del_\mu U, U^\dagger\del_\nu
U]^2 \nonumber\\
&&+{\mbox Tr}(M(U +U^\dagger-2)) +\cdots\label{skyrmeL}
 \ee
where $f$ is a constant related to the pion decay constant
$f_\pi\sim 93$ MeV, $e$ is an unknown constant, $U\in SU(3)$ is
the chiral field, $M$ is the three-flavor mass matrix and the
ellipsis ignored in the standard treatments stands for higher
dimension terms involving derivatives and mass matrices etc. The
skyrmion given by the Lagrangian (\ref{skyrmeL}) is then quantized
by what is called ``rigid rotor" method, that is,
collective-quantized with all flavors put on the same footing.

In describing baryons as chiral solitons, regardless of whether it
involves only chiral fields or a hybrid of chiral fields and quark
fields, there are two immediate issues in this approach. One is
how good the chiral Lagranigian used is as an effective theory of
QCD from which baryons are built and the other is how reliable the
quantization of the soliton is.  Some of these issues are
discussed in the monograph~\cite{NRZ}. Let me address the first of
these issues in this subsection relegating the second to the next
subsection.

The skyrmion is a description of baryons valid in the $N_c$ limit,
so we wish to have an effective Lagrangian that is close to QCD in
that limit. Let us look at (\ref{skyrmeL}). The first term of
(\ref{skyrmeL}) is the term that encodes the known current
algebras, so it is a highly trustful lowest-dimension term in
effective field theory. However the second term, so-called Skyrme
term, is one of several quartic terms that one can write down on
the basis of general dimensional and symmetry arguments. Given our
almost complete ignorance of QCD at low-energy nonperturbative
regime, however, it has not been feasible to ``derive" directly
from QCD terms other than the current algebra term. It is
plausible that a quartic term of the Skyrme-type arises from
heavier meson degrees of freedom of a given quantum number when
one integrates out all higher energy degrees of freedom other than
the pseudo-Goldstone bosons but there is no reason why other
vector mesons such as $a_1$ ... and heavy scalar mesons etc.
should not give rise to similar -- but different forms of --
four-derivative terms. In fact there is no theoretical reason
whatsoever why one would wind up with only one term like the
Skyrme term if one is ``deriving" them from QCD {\it per se}. As I
will mention below, an unexpected support for this Skyrme-type
term comes from string theory: Using ``AdS/QCD holographic
duality," QCD can be expressed as a compactified theory of
five-dimensional Yang-Mills gauge theory
~\cite{sakai-sugimoto,hill-zachos} from which at low energy the
Skyrme term arises with a coefficient fixed in terms of the string
variables in addition to a tower of vector mesons populating up to
a given cutoff scale. This term is there whether or not one takes
explicitly into account the low-lying vector mesons $\rho$,
$\omega$, $a_1$, etc. Given by the bulk sector in addition to the
tower of vector gauge fields, it must represent shorter distance
scales than the tower of vector mesons.\footnote{The presence of
the short-distance Skyrme term in addition to the tower of vector
mesons tells us that the Skyrme term should be present in models
where the lowest members of the tower, $\rho$ and/or $\omega$, are
used to stabilize the soliton as in ~\cite{AN,prv-omega}. This
short-distance term must be the Skyrme term that figures in the
description of the monopole catalysis of the nucleon decay given
in terms of the unwinding of the skyrmion~\cite{rubakov}. It would
also resolve the problem of the unrealistic $\omega$ repulsion in
dense medium seen in~\cite{prv-omega} where an $\omega$ with a
finite mass was used for stabilization.}

In most of the work on skyrmions (with a few exceptions), the
vector meson degrees of freedom have been ignored or integrated
out. In \cite{scoccolaetal}, the role of vector mesons in the
hyperon structure was first investigated and it was noted that
while the explicit account in the skyrmion structure of the
lowest-lying nonet vector mesons definitely improves the
spectroscopy for the $S<0$ baryons but there was no qualitative
influence. It will turn out however that the vector mesons in
hidden local gauge fields can make a drastic effect on $S>0$
baryons, i.e., pentaquarks. This will be seen to be closely linked
to hidden local symmetry.

It has been debated in the literature as to whether the skyrmion
description of pentaquark baryons can be taken seriously as
modelling QCD. It has even been suggested that it can be
invalidated by experiments. I would think that this debate will be
settled if indeed the AdS/QCD duality is correct. As mentioned,
the duality means that the skyrmion description for baryons is a
holographic dual to the quark descriptions, so they describe the
same physics. If there is any difference, it will only be in the
approximations one makes in doing the calculations, not in the
principle. A similar point of view is voiced by
Diakonov~\cite{diakonov} from a different starting point.
\subsection{Large $N_c$}
The next issue in the skyrmion description of the $\thetap$ has to
do with how one quantizes the skyrmion in consistency with the
$N_c$ counting. The $N_c$ dependence of the baryon mass can be
written as
 \be
M=A N_c + B + C/N_c +\cdots\label{mass}
 \ee
where $A$, $B$, $C$ ... are $N_c$-independent coefficients. The
$A$ term is the classical soliton mass common to all baryons and
the $C$ term is the rotational contribution that comes from
collective quantization that splits a state with spin $J=3/2,...,
N_c/3$ from the ground state with $J=1/2$. The constant $B$ term
contains several contributions among which the Casimir energy is
the most important one. The Casimir calculation in baryon
structure turns out to be horrendously difficult. There have been
several courageous attempts (see for reference, ~\cite{NRZ}), but
no reliable result has been obtained yet. In most of the the
skyrmion work available in the literature, one simply ignores the
Casimir energy and uses (\ref{mass}) taking the constants $f$ and
$e$ as arbitrary parameters fit to a set of low-energy data. As
argued in \cite{NRZ}, this is not the right way of doing things.
An indication that this procedure cannot be reliable is that the
fit value of $f$ is drastically different from what it is in
nature, viz,  the pion decay constant $f_\pi\approx 93$ MeV.
Although this fitting procedure may make sense in some cases, it
cannot be applied generally since the ${\calO}(1)$ terms can in
general depend on the quantum numbers involved (e.g.,
strangeness). Furthermore the Casimir calculations performed so
far indicate that the ${\calO}(1)$ term can be very big, though
subleading in $N_c$, and attractive. The correct procedure should
be to take properly into account the Casimir energy and then use
the physical $f_\pi$ for the skyrmion properties.

The formula (\ref{mass}) applies equally to the non-exotic baryons
as well as to the exotic baryons with the coefficients $B$ and $C$
tracking the quantum numbers of the baryons. It has been pointed
out by several authors~\cite{klebanov,cohen} that the rigid rotor
expression for the mass formula (\ref{mass}) for the exotic
channel is inconsistent with the $N_c$ counting. The argument goes
as follows. While the mass splitting between the multiplets in the
non-exotic channel is given by ${\calO} (1/N_c)$ term -- there is
no ${\calO} (1)$ contribution, the splitting from the lowest
(non-exotic) multiplet in the exotic channel has a term $\sim
N_c/\Phi\sim {\calO}(1)$ (where $\Phi$ ia a moment of inertia in
the strangeness direction) originating from the rigid rotation.
But the rigid rotor method is believed to miss out some
potentially important terms of the same order. For instance, in
addition to the Casimir energy, there can be possible
``vibrational modes" in the strangeness direction (as in the
bound-state quantization, see below) that are not accounted for.
This inconsistency in ${\calO}(1)$ has been invoked~\cite{cohen}
to cast doubt on the prediction despite the amazing agreement with
the experiment.

An alternative approach discussed by the authors in
\cite{klebanov,cohen} avoids this inconsistency in $N_c$ counting.
The method consists of fluctuating in the strangeness +1 direction
in the background of an $SU(2)$ soliton, with the Wess-Zumino term
playing the role of a magnetic field. The question is: What
happens to a $K^+$ in the presence of the soliton? This approach
was first developed for the $S<0$ baryons~\cite{CK} where $K^-$'s
were found to be bound to the soliton to give the known hyperons.
In fact it is this binding approach that one should resort to when
one is considering heavy-light-quark baryons since that is the
natural way to incorporate heavy-quark symmetry into a chiral
symmetric Lagrangian. Here the kaon can be thought of as a
vibration -- which is of ${\calO}(1)$ -- as contrasted to the
rotation -- which is typically of ${\calO}(1/N_c)$. Since the kaon
vibrational energy is of the same order in $N_c$ as the Casimir
energy, in principle both should be treated on the same footing.
Due to the difficulty with the Casimir calculation, a consistent
calculation of this sort has not been done. I would think that
such a calculation will ultimately be needed to understand the
$\thetap$ structure as I will mention at the end.

The crucial element in the bound state picture is the Wess-Zumino
term responsible for the five pseudoscalar coupling, $K^2 \pi^3$.
In the description with (\ref{skyrmeL}) supplemented with the
topological Wess-Zumino term, this coupling -- the strength of
which is fixed by topology -- is attractive for the $S<0$ channel
accounting for the binding. However for the $S>0$ channel, the
sign change makes the Wess-Zumino term repulsive. Therefore with
the standard parameters that describe the non-exotic baryons,
there cannot be binding nor resonance with a small
width~\cite{klebanov}. It was found that if the kaon mass were
much bigger (i.e., stronger $SU(3)$ breaking) or the Wess-Zumino
term were smaller by a factor of $\sim 2$, a $K^+$ could be
quasi-bound or produce a narrow-width resonance.

This model as it stands therefore cannot accommodate a
narrow-width resonance of the $\thetap$ quantum number.

This result may also augur on the validity of the rigid rotor
description~\cite{klebanov}. In the limit that the kaon mass goes
to zero, the rigid rotor zero mode is recovered smoothly as
$m_K\rightarrow 0$ for the $S<0$ channel. The bound-state and the
rigid rotor models agree in the chiral limit. On the contrary, for
the $S>0$ channel, the bound-state model has no solution that goes
over to the zero mode of the rigid rotor model. This result has
been confirmed by the calculation with hidden local symmetry
theory to be described below~\cite{PRM} that does support a bound
state for a reasonable parameter of the HLS Lagrangian.

Several questions are raised at this point. Is it that the
skyrmion model as a whole, independently of how it is quantized,
does not accommodate the $\thetap$ baryon? Does this mean that the
rigid rotor result which gave the uncanny prediction is
accidental? Some of these questions will be addressed below.
\section{Vector mesons as hidden gauge fields}
When the light-quark vector mesons are included as relevant
degrees of freedom, there can be a dramatic change in the
structure of the $S>0$ baryons although they leave more or less
unaffected the $S<0$ baryons. In this respect, treating the vector
mesons as gauge fields has a power that is not available in
non-gauge invariant approaches. I shall discuss why local gauge
invariance possesses such power and how such local gauge
invariance can arise from first principles.
\subsection{Hidden local symmetry}
To bring out a general idea, let me consider, following
Ref.\cite{georgi}, a massive vector field Lagrangian
 \ba
{\calL}= - \frac{1}{2g_2^2}{\mbox Tr}F_{2\mu\nu}^2 +
\frac{f^2}{4}{\mbox Tr} A_{2\mu}^2 +\cdots.\label{massiveL}
 \ea
which is a Lagrangian describing a massive vector excitation with
the mass
 \ba
 m_{A_2}=f g_2.
 \ea
This can be gotten by taking $g_1=0$ from
 \ba
{\calL}&=&-\frac{1}{2g_1^2}{\mbox Tr}F_{1\mu\nu}^2 -
\frac{1}{2g_2^2}{\mbox Tr}F_{2\mu\nu}^2 + \frac{f^2}{4}{\mbox Tr}
(A_{1\mu}-A_{2\mu})^2\nonumber\\
&& +\cdots.\label{gauge-fL}
 \ea
Being massive, the Lagrangian (\ref{massiveL}) or (\ref{gauge-fL})
has no gauge symmetry. But introducing a Goldstone scalar field
$U=e^{i\pi/f}$ which transforms as
 \ba
U\rightarrow g_2^{-1} U g_1
 \ea
one can write a gauge invariant-Lagrangian that has the same
low-energy physics as (\ref{gauge-fL}),
 \ba
{\calL}=-\frac{1}{2g_1^2}{\mbox Tr}F_{1\mu\nu}^2 -
\frac{1}{2g_2^2}{\mbox Tr}F_{2\mu\nu}^2 + \frac{f^2}{4}{\mbox Tr}
|D_\mu U|^2 +\cdots
 \ea
with the covariant derivative defined by
 \ba
D_\mu U=\del_\mu + iA_{1\mu} U - i UA_{2\mu}.\label{gauge-iL}
 \ea
This Lagrangian is invariant under the gauge symmetry
$SU(n)_1\times SU(n)_2$. To get to (\ref{gauge-fL}), it suffices
to gauge fix $U$ to the unitary gauge, $U=1$.

So what is the big deal with the gauge invariance?

Let us suppose that the Lagrangian (\ref{massiveL}) is only a part
of an effective theory to be embedded into a more complicated
system. At higher orders that should come in as one goes up in
scale, there would be terms of the form
 \ba
\frac{1}{16\pi^2}{\mbox Tr}A^4, \ \ \frac{1}{16\pi^2}{\mbox
Tr}(\del A)^2, \cdots\label{higherderivative}
 \ea
Such terms are naturally generated when one calculates
higher-order terms by doing loop calculations which requires that
additional equivalent terms be added to renormalize the loop
terms. Without any systematic ways of accounting for higher order
terms, it will require an arduous book-keeping to correctly put
all terms of the same order. Local gauge symmetry can do this
systematically with little effort by replacing
(\ref{higherderivative}) by the covariant derivatives
 \ba
\frac{1}{16\pi^2}{\mbox Tr}|D_\mu U|^4, \ \
\frac{1}{16\pi^2}{\mbox Tr} |D^2 U|^2,
\cdots\label{covariantdervative}
 \ea
This allows a systematic account of higher-order terms to any
order. This is important in the case where we have not only the
Goldstone bosons, e.g., pions, as physical particles but also the
vector fields whose mass can be $made$ light as in hot and/or
dense matter~\cite{BR-PR96}.

Another advantage is that with the Goldstone bosons explicitly
treated instead of being ``eaten up," the theory of the massive
vector bosons remains sensible in the effective theory sense up to
the cutoff $\sim 4\pi m_A/g\sim 4\pi f$. Above that scale, a
microscopic theory  has to take over (that is, be ``ultraviolet
completed"). In our case, the ultraviolet completion is done to
QCD.
\subsection{Hidden local symmetry \`a la Harada and Yamawaki}
There is a growing evidence that QCD at low energy can be
described as a chiral $SU(N_f)\times SU(N_f)$ symmetric theory
with a tower of vector mesons, scalars and Goldstone bosons, all
coupled gauge invariantly. The structure I will use here for the
pentaquatk problem is the one where only the lowest multiplet of
the vectors $\rho_\mu$ and pions $\pi$ are taken into account,
with scalars and higher towers of the vectors ``integrated out."
The Lagrangian will be taken in the
form~\cite{bandoetal,HYPR}~\footnote{I am using a different
notation from that of Harada and Yamawaki to generalize the
Lagrangian as discussed below. One can identify
$\Sigma_L=\xi_L^\dagger$ and $\Sigma_R=\xi_R$. Note that I use
here $\kappa$ in place of $a$, with $\kappa=0$ corresponding to
$a=1$ which will play a prominent role in what follows. In what
follows, $a$ will be used instead of $\kappa$.}
 \ba
{\cal L}&=&\frac{F_\pi^2}{4}{\mbox Tr}\{|D_\mu\Sigma_L|^2
+|D_\mu\Sigma_R|^2+ \kappa |D_\mu (\Sigma_L\Sigma_R)|^2\}
\nonumber\\
&& - \frac{1}{2g^2} \, \mbox{Tr} \left[ \rho_{\mu\nu}
\rho^{\mu\nu} \right] + \cdots\label{hls}
 \ea
with $\Sigma_{L/R}$ transforming under chiral $SU(N)_L\times
SU(N)_R$ as
 \ba
\Sigma_L \rightarrow L\Sigma_L h^\dagger (x),\ \ \Sigma_R
\rightarrow h (x)\Sigma_R R^\dagger
 \ea
with $L\in SU(N_f)_L$, $R\in SU(N_F)_R$ and $h(x)\in
SU(N_f)_{L+R}$. In an arbitrary gauge, $\Sigma_{L,R}$ can be
parameterized as
 \ba
\Sigma_{L/R}=e^{i\pi/F_\pi} e^{\pm i\sigma/F_\sigma.}
 \ea
With the definition
 \ba
 a=(F_\sigma/F_\pi)^2,\label{a}
 \ea
 $\kappa$ is given
by
 \ba
\kappa=\frac{a-1}{a+1}.
 \ea

It is interesting to note that $\kappa=0$ or $a=1$ has a special
feature. The Lagrangian becomes ``local" with no mixing of the L
and R chiral fields. One refers to this structure as ``theory
space locality"~\cite{DC-georgi}. If one preserves this theory
space locality, one can then make a chain of gauge fields $A_k$
sandwiched by the ``link" fields $\Sigma_k$ and $\Sigma_{k+1}$.
This gives what is referred to in the literature as ``open moose"
diagram with the Lagrangian~\cite{son-stephanov1}
  \ba
{\it L}=\sum_{j=1}^{K+1}\frac{f_j^2}{4}{\mbox Tr}|D\mu\Sigma^j|^2
\ - \sum_{j=1}^{K}\frac{1}{2g_j^2} \, \mbox{Tr} \left[
F^j_{\mu\nu} F^{j\mu\nu} \right]\ . \label{hlsK}
 \ea
Here we have replaced the ``pion decay constant" $F_\pi$ by a
generalized constant $f^j$ corresponding to the $j$-th link field
and $\rho_{\mu\nu}$ by $F^j_{\mu\nu}$ to accommodate $K\geq 1$
gauge fields. The covariant derivatives are of the form $D_\mu
\Sigma^1=\del_\mu\Sigma^1 +i\Sigma^1 A_\mu^1$, $\ D_\mu
\Sigma^{K+1}=\del_\mu\Sigma^{K+1} -iA_\mu^K\Sigma^{K+1}$ and
$D_\mu \Sigma^k =\del_\mu\Sigma^k -iA_\mu^k\Sigma^K +i\Sigma^k
A_\mu^1$ with $1<k<K$. If we let $K\rightarrow \infty$, then
(\ref{hlsK}) can be interpreted as a Lagrangian for
five-dimensional gauge theory with the fifth dimension put on
lattice. Indeed making it continuous leads to a gauge invariant
theory given in a generally curved space (that is, with a non-flat
metric) in five dimensions. Physically what that means is that by
putting in the infinite tower of vector mesons with local gauge
invariance, one can ``deconstruct" the fifth dimension which
represents the energy
scale~\cite{DC-georgi,DC-hill,son-stephanov1}.
\subsubsection{AdS/QCD}
In \cite{son-stephanov2}, Erlich et al interpret the dimensionally
deconstructed five-dimensional gauge theory implemented with
certain scalar fields as the bulk theory in the (5-D) AdS space
which via holography correspond to low-energy QCD in four
dimensions on the boundary. They obtain, using the strategy of
AdS/CFT duality, quite a reasonable low-energy hadron dynamics
with three bulk parameters fit to the pion and $\rho$ masses and
the pion decay constant. As already mentioned above and reported
in this meeting by Sugimoto, Sakai and
Sugimoto~\cite{sakai-sugimoto} go the other way. They insert $N_f$
pairs of flavor D8 branes as probes into the AdS sector with a
background of $N_c$ color D4 branes and assuming $N_c\gg N_f$,
obtain five-dimensional gauge theory which when compactified to
four dimensions, gives HLS theory with a tower of vector mesons,
with the parameters of the theory fixed by the bulk parameters.
The results agree fairly well with experiments. I cannot give a
full justice to this remarkable feat as the calculation requires
involved arguments in a language not entirely familiar to me. But
what is most noteworthy is that the scheme gives in four
dimensions precisely the hidden local symmetry structure of Bando
et al~\cite{bandoetal} and Harada/Yamawaki~\cite{HYPR} including
the gauged Wess-Zumino term and $furthermore$ fixes the Skyrme
quartic term uniquely in terms of the bulk parameters. {\it An
important consequence of this is that baryons must emerge in
AdS/QCD as skyrmions in the hidden local symmetry theory.}

What appears to be potentially important for the pentaquark
structure I will describe below is that the Sakai-Sugimoto model
predicts what corresponds to $a$, Eq.(\ref{a}), in the limit of
large $N_c$,
 \ba
a\approx 1.3.
 \ea
It is close to what was obtained at the matching point by Harada
and Yamawaki in their best-fit analysis~\cite{HYPR} and is closer
to 1 than to 2 required by the vector dominance as will be further
elaborated below.

\subsection{The vector manifestation}
Little is known of the theories with $K>1$ in (\ref{hlsK})
although there are some developments that indicate that the
dimensionally deconstructed theory is close to
nature~\cite{son-stephanov2}. I shall therefore focus on the
Harada-Yamawaki HLS theory (\ref{hls}) which has been extensively
analyzed and fairly well understood~\cite{HYPR}. We will be
dealing with the theory that should work up to the cutoff given by
$\Lambda_M \sim 4\pi m_\rho/g\sim 4\pi f_\pi\sim 1$ GeV; we can
think of having integrated out all except the pions and the nonet
vector mesons (assuming flavor $U(N_f)$ symmetry) generically
denoted by $\rho_\mu$. The axial vectors are not considered
explicitly.

The basic premise in HY theory is that the strong interaction
dynamics is governed by the hadronic degrees of freedom picked for
consideration, i.e., the pions and vector mesons, below the cutoff
$\Lambda_M$ and by the QCD degrees of freedom, quarks and gluons,
above the cutoff. It is assumed that there is a region in between,
say, $\Delta$, in which the two regimes overlap. It is not obvious
that such a region can be picked reliably without ambiguity but
this is the best one can do if the effective theory is to be
operative up to around $\Lambda_M$ beyond which QCD has to take
over. HY do this by matching at $\Lambda_M \sim 1.1$ GeV the two
descriptions in terms of the vector and axial correlators defined
by
\begin{eqnarray}
i \int d^4 x e^{i p x} &&\left\langle 0 \left\vert T\, J_{5\mu}^a
(x) J_{5\nu}^b (0) \right\vert 0 \right\rangle \nonumber\\
 &&=
\delta^{ab} \left( p_\mu p_\nu - g_{\mu\nu} p^2 \right) \Pi_A
(p^2) \ ,
\nonumber\\
i \int d^4 x e^{iqx} && \left\langle 0 \left\vert T\, J_{\mu}^a
(x) J_{\nu}^b (0) \right\vert 0 \right\rangle \nonumber\\
 && =
\delta^{ab} \left( p_\mu p_\nu - g_{\mu\nu} p^2 \right) \Pi_V
(p^2) \ , \label{A V correlators 4}
\end{eqnarray}
where $Q^2 = -p^2$. In HLS these two-point functions at the
matching point $\Lambda_M$ are completely described by tree
contributions with ${\cal O}(p^4)$ terms included when the
momentum is around the matching scale, $Q^2 \sim \Lambda_M^2$. In
the QCD sector they are given by OPE expressed in terms of the
running (color) gauge coupling $\alpha_s$, the quark condensate
$\la\bar{q}q\ra$, the gluon condensate $\la G_{\mu\nu}^2\ra$ and
combinations thereof at the given scale $\Lambda_M$. The matching
allows the parameters $F_\pi$, $F_\sigma$ or $a$ and the hidden
gauge coupling $g$ to be expressed in terms of the QCD quantities
that are in principle calculable at a given scale. This completely
defines the $bare$ Lagrangian for HLS theory. In this theory, the
vector mesons can be treated as ``light" in the chiral counting
scheme. Thanks to gauge invariance, in the large $N_c$ limit at
which the quantity $m_\rho/4\pi F_\pi$ with a fixed vector meson
mass is suppressed, higher orders in chiral perturbation expansion
can be systematically formulated, an advantage which is not shared
by non-gauge invariant or gauge fixed approaches.

The most important outcome of the systematic analysis of the
theory at loop orders~\cite{HYPR} is that there is one unique
fixed point which is consistent with QCD. In general there are
multitudes of fixed points at one-loop order. However constraining
the RGE group flow to QCD, the unique fixed point turns out to be
at $g=0$, $a=1$ and $f_\pi=0$. This is called the ``vector
manifestation (VM) fixed point." Whenever appropriate, I shall
refer to the HLS theory with the vector manifestation fixed point
as HLS/VM. This fixed point is obtained in the chiral limit but is
believed to be relevant and useful even when quark masses are
present. I shall therefore talk about the VM fixed without regard
to the explicit chiral symmetry breaking.

The VM fixed point is reached when (in the chiral limit) chiral
symmetry changes from Goldstone mode to Wigner mode or vice versa.
It does not depend upon how the phase change is driven. It could
be by temperature or density or the number of flavors. The VM
implies that as the hadronic system approaches the critical point
${\cal P}$,
 \ba
a \rightarrow 1,\ \ g \rightarrow 0,\ \ f_\pi \rightarrow 0
 \ea
so that the hidden gauge bosons become massless
 \ba
m_\rho\sim \sqrt{a} g f_\pi\rightarrow 0.
 \ea
Here $m_\rho$ is the pole mass but the parametric mass also goes
to zero at that point. Matching with QCD allows one to relate the
scaling of the parameters to the scaling of QCD variables. Thus as
one approaches ${\cal P}$,
 \ba
m_\rho \sim g \sim \la\bar{q}q\ra\rightarrow 0.
 \ea
Applied to $N_f=2$, this implies that near ${\cal P}$, $all$
light-quark hadron masses other than the Goldstone bosons should
scale as
 \ba
M^\star/M \sim (\la\bar{q}q\ra^\star/\la\bar{q}q\ra)^n\label{BR}
 \ea
where the $\star$ stands for in-medium -- temperature or density
-- and $n>0$ a power that can depend upon the detail content of
the theory. One can think of baryons as skyrmions in the theory or
one can introduce quasiquarks (or constituent quarks) considered
to be relevant near the critical point and express the scaling in
terms of that of the quasiquarks. This scaling is precisely what
was proposed in 1991~\cite{BR91}.
\section{HLS/VM and pentaquark structure}
\subsection{Evidence for $a\approx 1$ in nature}
Even though a hadronic system is not at the VM fixed point, with
$g\neq 0$ and $f_\pi\neq 0$, it appears to be a good approximation
to set
 \ba
a\approx 1
 \ea
and make corrections in powers of $\delta=(a-1)$. Let me list a
few cases known up to now where this holds.
 \ben
 \item {\boldmath $\Delta m_\pi^2=m_{\pi^+}^2-m_{\pi^0}^2$}:
\vskip 0.3cm In computing the pion mass difference in HY's
HLS/VM~\cite{pionmass}, one gauges the $U(1)$ subgroup of the
chiral symmetry in HLS to introduce the EM field and takes into
account the mixing of the $U(1)$ gauge field and $\rho^0$ of the
$SU(2)$ hidden gauge field. Then the one-loop EM contribution to
the mass difference comes out to be
 \ba
\Delta m_\pi^2|_{loop} =
\frac{\alpha_{em}}{4\pi}&&[(1-a)\Lambda^2+3a M_\rho^2
\ln\Lambda^2\nonumber\\
&&+\cdots].\label{photonloop}
 \ea
Here $\Lambda$ is the cutoff introduced to regularize the
divergent integreal. In HLS/VM, it is identified as the matching
scale $\Lambda=\Lambda_M$. Beyond $\Lambda_M$, the theory has to
be ultraviolet completed and this is done by matching to QCD via
the current correlators. The result is~\cite{pionmass}
 \ba
\Delta m_\pi^2|_{>\Lambda_M}=\frac{8}{3}
\frac{\alpha_{em}\alpha_s\la\bar{q}q\ra^2}{F_\pi^2
(0)\Lambda_M^2}.
 \ea
Assume that the number of flavors $N_f$ is sufficiently ``large"
so that we can consider $a$ to be near its fixed point, $a=1$.
Then ignoring higher orders in $\delta$, we can set $a=1$ in
(\ref{photonloop}) and get the total splitting as
 \ba
\Delta
m_\pi^2&=&(3\alpha_{em}/4\pi)M_\rho^2\ln(\Lambda_M^2/m_\rho^2)\nonumber\\
 && +
\Delta m_\pi^2|_{>\Lambda_M}.
 \ea
The result $\Delta m_\pi^2|_{HLS/VM}=1223\pm 263$ MeV$^2$ obtained
using the value $a=1$ and other parameters fixed in a best-fit to
hadronic structure~\cite{HYPR} is in good agreement with the
experimental value $1261$ MeV$^2$.

An interesting observation here is that the $a=1$ leads to the
theory space locality eliminating the ``quadratic divergence" in
(\ref{photonloop}). This cancelation which is effectuated in the
standard current algebra approach with the Weinberg sum rules with
$\pi$, $\rho$ and $a_1$ is the analog to what happens in the
little Higgs phenomenon in the hierarchy resolution in the
Standard Model.

If instead of fixing $a=1$ one uses Harada-Yamawaki's global-fit
value $a\approx 1.3$, one obtains a somewhat less good but still
satisfactory agreement with the experiment for the pion mass
splitting. \vskip 0.3cm
\item {\bf The chiral doubling of heavy-light hadrons}:
\vskip 0.3cm
 Another recent indication that  $a$ is close
to 1 in nature is given by the splitting between the chiral
doublers of heavy-light mesons $H\equiv (Q\bar{q})$ of open heavy
quark $Q=c, b, ...$ and light antiquark $\bar{q}=\bar{u}, \bar{d},
\bar{s}$. It was predicted in 1992 on the basis of sigma models
combining chiral symmetry and heavy quark symmetry and published
in 1993~\cite{nrz-HL,bh-HL} that the mass splitting $\Delta
M=m_{H(0^+,1^+)}-m_{H(0^-,1^-)}$ is given approximately by the
constituent (quasiquark) mass $\Sigma\approx m_N/3\approx 313$
MeV. This prediction was confirmed recently by the BarBar and
CLEOII collaborations~\cite{babar,cleoII}. One unsatisfactory
feature of the old prediction was that it was given in terms of
the constituent quark mass which is not well defined physically.
This defect can be lifted in an approach based on
HLS/VM~\cite{hrs-babar}.

The basic assumption made in \cite{hrs-babar} is that the VM is a
good starting point as far as chiral symmetry is concerned. So one
starts with HLS with the VM fixed point parameters and then makes
corrections for deviation from the VM as one departs from the
chiral transition point. In doing this, one takes $a=1$ and
computes the deviations in other parameters, namely, $g\neq 0$ and
$f_\pi\neq 0$. Because of the matching to QCD, one can then
express $\Delta M$ in terms of the QCD variables, in particular
the quark condensate $\la\bar{q}q\ra$. The result has the simple
form
 \ba
\Delta M=C_{quantum}\Delta M_{bare}
 \ea
where
 \ba
\Delta M_{bare}=-\frac 23 \frac{\la\bar{q}q\ra}{F_H^2}
 \ea
with $F_H$ the $H$-decay constant is the contribution given by
matching with QCD at the matching scale and $C_{quantum}$ is the
quantum loop correction (given, as it turned out, entirely by the
$\rho$ loop) calculated by RGE. For the $D$ mesons, the quantum
correction comes out to be $C_{quantum}=1.6$. With this and the
available values for $F_D$ and the quark condensate, one
reproduces the constituent quark mass $m_N/3\approx 313$ MeV. Thus
the VM is not far from the real world of the chiral doublers. This
prediction is a lot more clear-cut than that of
\cite{nrz-HL,bh-HL} since the quark condensate is a well-defined
quantity for a given scale. \vskip 0.3cm
 \item {\bf Electromagnetic form factors of the proton}
\vskip 0.3cm Although $a$ at the matching scale is near 1, quantum
corrections (i.e., $1/N_c$ corrections) {\it in the mesonic
sector} bring it to $a\approx 2$ at the scale corresponding to the
pion on-shell. $a=2$ gives rise to the vector dominance for the
pionic EM form factor. Now the situation with the nucleon is a
different matter. There is no systematic study of quantum
($1/N_c$) corrections to $a$ for the baryon sector, so we have no
theoretical information on how good the VD is for nucleons or
baryons in general. But there are empirical evidences that the VD
with $a=2$ breaks down for the nucleon form factors. In fact there
is a strong indication that for the nucleons, $a\approx 1$ is a
good approximation, as suggested a long time ago\cite{IJL,brw}
based on phenomenological considerations and confirmed by new data
from JLab.

How the violation of VD in the nucleon form factors comes about
was first explained years ago in terms of a chiral bag
model~\cite{NRZ,RGB83}. In chiral bag model, the electromagnetic
form factors have two sources: One is the contribution from the
weakly interacting free quarks confined within a bag and the other
is from the pion cloud outside of the bag, with the two regions
connected by chiral-invariant boundary conditions. The partition
into two regions is characterized by the chiral angle $\theta$. At
the ``magic angle" $\theta=\pi/2$, one has one unit of baryon
charge partitioned equally inside and outside of the bag. The
outside pion cloud is assumed to respond to the electromagnetic
field via the vector ($\rho$, $\omega$) mesons in accord with VD
whereas the quarks inside are to respond as a compact core of size
$\lsim 0.3$ fm, i.e. the bag at the magic angle. This description
can be translated into the modern version of effective field
theory by constructing baryons as skyrmions in
HLS~\cite{BR-PR-dd}. In HLS, the photon will couple to the pion
cloud weighed by $a/2$ and to the soliton core by $(1-a/2)$
 \ba
\delta {\calL}&=&-2eag F_\pi^2 A_{em}^\mu{\mbox Tr}[\rho_\mu
Q]\nonumber\\
&&+2ie(1-a/2) A_{em}^\mu {\mbox Tr}[V_\mu Q]
 \ea
where $A_{em}^\mu$ is the photon field, $Q=\frac 13 {\mbox{diag}}
(2,-1,-1)$ is the quark charge matrix and $V^\mu$ is the matter
vector current. At $a=1$, the coupling is half-and-half. It was
argued at the time~\cite{brw} that this described best the
available data on the nucleon form factor. This picture is very
well supported by the recent JLab experiment on the form factor
ratio $\mu_p G_E/G_M$ of the proton measured to large momentum
transfers~\cite{iachello}.

As suggested in \cite{BR-PR-dd}, the $a$ approaching 1 may be
generic in the presence of matter. In fact. a careful analysis of
the RGE flow of the HLS theory with the VM shows that the vector
dominance with $a=2$ at the pion on-shell is an ``accident" on an
unstable trajectory, so $a$ can be driven away by small external
perturbation from the VD point~\cite{HY:vd}. This was confirmed
for hadronic matter in heat bath~\cite{sasaki-T}: The VD is
predicted to be maximally violated as temperature drives $a$ to 1
quickly. The claim in \cite{BR-PR-dd} is that in hadronic matter,
$a$ is driven to near 1. This has not yet been checked by explicit
calculations in HLS for baryonic matter. It could however be
checked by experiments. I propose to generalize this: That {\it
any system that involves baryons, whether in isolation (e.g.,
nucleon, pentaquark etc.) or in many-body aggregation (nuclei,
nuclear matter etc.), would have $a\approx 1$}. I will apply this
proposition to the pentaquark structure.

 \een
\subsection{Bound pentaquarks for $a$ near 1}
I am now in a position to discuss what happens to pentaquarks in
HLS/VM~\cite{PRM}. The main thesis in my discussion is that {\it
the kaon-skyrmion interaction takes place in a baryonic matter and
hence the parameter $a$ will be driven away from the VD value of 2
to near 1 of VM}.

\begin{figure}[tb]
\centerline{\epsfig{file=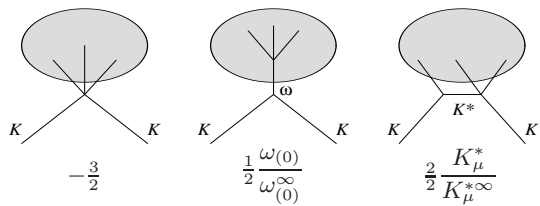,width=1.8cm,height=7cm,angle=270}}
\setlength{\unitlength}{1mm}
\begin{picture}(80,8)
\put(15,4){\makebox(0,0)[c]{$-\frac32$}}
\put(40,4){\makebox(0,0)[c]{$\frac12\displaystyle
          \frac{\omega_{(0)}}{\omega^\infty_{(0)}}$}}
\put(65,4){\makebox(0,0)[c]{$\frac22\displaystyle
          \frac{K^*_\mu}{K^{*\infty}_\mu}$}}
\end{picture}
\caption{\small Schematic illustration of the WZ-like terms due to
vector mesons in the Lagrangian. The first figure corresponds to
the topological WZ term and the other two WZ-like terms generated
by the vector mesons $\omega$ and $K^\star$. The baryon number
density is given by the three pion lines connected to the shaded
area which represents the soliton. \label{WZ:illust}}
\end{figure}

Let me begin by observing that qualitatively speaking, when
vector-meson degrees of freedom are present, there are two
important changes that occur in the interaction of the kaon with
the $SU(2)$ soliton. Firstly the strange vector meson $K^\star$
couples to the soliton as the kaon does. Thus the fluctuation in
the strangeness direction involves a coupled channel problem. The
obvious effect is the level repulsion between $K$ and $K^\star$. A
much subtler effect is that the presence of the ``light" mass
vector mesons, $\omega$ and $K^\star$, brings in a solitonic
matter additional terms that behave like the Wess-Zumino term in
the coupling of $K^2\pi^3$. This may be stated as saying that the
``effective magnetic field" played by the Wess-Zumino term is
modified. This is illustrated in Fig.\ref{WZ:illust}.
\begin{figure}[h]
\centerline{\epsfig{file=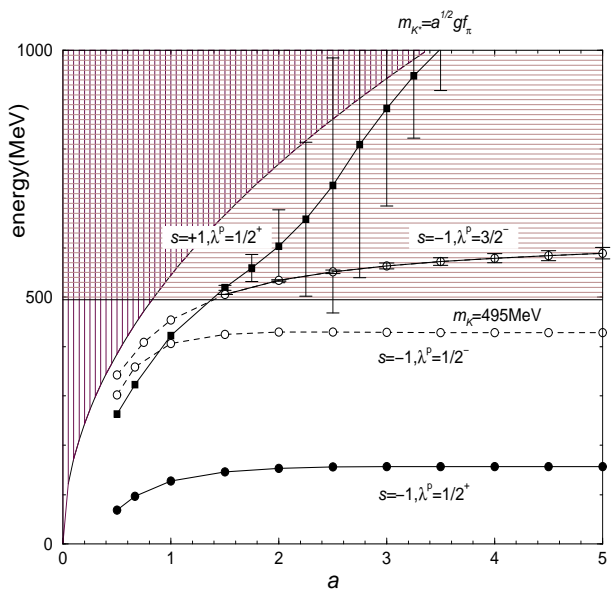,width=8cm,height=8cm,angle=270}}
\caption{The eigenenegies of $S=\pm 1$,
$\lambda^P=\frac12^\pm,\frac32^-$ states obtained for various $a$
values. The width of the resonance in the $K^+$-skyrmion channel
is given by the error bar. \label{e-vs-a}}
\end{figure}
Now with hidden local symmetry, one can show~\cite{mrs} that the
vector meson mediated graphs cancel each other in the limit of the
infinite vector meson mass, leaving the topological WZ term which
is effectively repulsive for the $K^+$ channel. However for finite
vector meson masses, there are non-local contributions to the
total WZ-like term which depending upon the value of $a$, can be
effectively repulsive or attractive in conjuction with the
$K^\star-K$ level repulsion.

A full spectroscopy of the pentaquark states requires collective
quantization and the proper account of all ${\calO} (1)$ terms and
$1/N_c$ corrections. Such a comprehensive calculation has not been
done yet. However to see qualitatively what happens to a $K^+$ in
the background of the soliton to order ${\calO} (N_c^0)$, it
suffices to study the coupled equations of motion for the $K$ and
$K^\star$ interacting with the soliton. For this purpose, we need
to take the large $N_c$ limit of the HLS Lagrangian given by the
$bare$ Lagrangian obtained at $\Lambda_M$ by the Wilsonian
matching. As noted, the $bare$ HLS Lagrangian obtained in a best
fit by HY~\cite{HYPR} gives $a\approx 1.3$ (and other parameters I
will not quote). For this Lagrangian, the equations of motion give
the $K^+$ bound by $\sim 5$ MeV. If one instead fixes $a=1$ at the
matching point and do the best fit for other two parameters, a
procedure which was found in \cite{HYPR} as acceptable as the
``best fit," the resulting results are in equally good agreement
with experiments.  The $K^+$ is also bound in this case, with the
binding energy $\sim 3$ MeV. Thus it can be concluded that {\it
for $a\lsim 1.3$, the $K^+$ is found to be definitely bound}
although by a small amount~\cite{MPR}\footnote{I would like to
acknowledge valuable help from Masayasu Harada and Dong-Pil Min
for these numerical estimates.}. To have a more general and
qualitative idea as to what is happening, the ``parameter" $a$ is
arbitrarily varied with all other parameters fixed to physical
(on-shell) values. This procedure is not quite justified for $a$
close to 1 since the parameters are correlated by the consistency
of the theory but it gives some idea as to how the $K^+$-soliton
depends on $a$ for $a>1$. The result~\cite{PRM} is shown in
Fig.\ref{e-vs-a}. What is given is the eigenenergy of the $K^\mp$
in the soliton background giving the system with $S=\mp 1$,
$J^\pi=\frac 12^\pm, \frac 32^-$. When the energy is less than the
free kaon mass $m_K=495$ MeV, the system is bound. When $a$ goes
to infinity, that goes over to the case studied in
\cite{klebanov}, namely, pseudoscalar-only theory.

We see that the $S=-1$ states are more or less insensitive to $a$
for a wide range of $a$~\footnote{This explains why Scoccola et
al~\cite{scoccolaetal} found the effect of the vector mesons in
the hyperon ($S<0$) spectroscopy to be undramatic. They failed to
look at the $S>0$ channel where the effect is a lot more
dramatic.} but the $S=+1$ state is strongly sensitive, with the
state bound for $a\lsim 1.3$ and unbound above. The widths of the
unbound states that show up as resonances (for $a\gsim 1.4$) are
given by the error bars. The error bars explode as the eigenenergy
increases~\footnote{For instance, at the observed mass difference
between the $\Theta^+$ and the proton, say, $\sim 100$ MeV,
corresponding to $a\approx 2$ for which the vector dominance
holds, the width can be $\sim 150$ MeV, clearly too big for the
``observed" $\thetap$.}.
\subsection{Confronting nature}
In order to confront nature, one would have to calculate the
Casimir contribution to the masses for the $\Theta^+$ as well as
the nucleon and quantize them including terms of ${\calO}(1/N_c)$.
The coupling of the quantum $\Theta^+$ to the proton-kaon
continuum will be at ${\calO}(1/N_c)$. In the absence of these
quantities, I can only make a few guesses. What is known is that
if naively collective-quantized with the Casimir term ignored, the
$\Theta^+$ state (near $a=1$) has a mass $m_{\Theta^+}$ lower than
the $KN$ threshold  $m_p + m_K$. The crucial question then is:
What happens when other corrections of ${\calO}(1)$ as well as
${\calO}(1/N_c)$ are included? Here are a couple of possibilities
that one can entertain~\cite{PRM,MPR}:
 \bitem
\item One possibility is that after the account of the corrections
mentioned above, a genuine bound state of the quantum numbers of
$\Theta^+$ hitherto unobserved experimentally remains below the
$KN$ threshold. In addition, there can be a bound
$K^\star$-soliton state in the $KN$ continuum weakly coupled to
the latter. This could be identified as the observed narrow
$\thetap$.
\item Another possibility is that higher-order
 $1/N_c$ corrections (starting with ${\calO}(1)$) are more attractive
for the nucleon than for the $\Theta^+$ such that the bound
``closed" channel state goes into the $KN$ continuum. In this
case, the observed peak could be generated as a ``Feshbach
resonance"~\cite{jain-jaffe} with a small width due to the weak
(${\calO}(1/N_c)$) $\Theta^+$-$K$-$N$ coupling. A possible
mechanism is that the baryonic matter of the soliton that dials
$a$ which in turn controls the WZ-like term could play, in the
presence of $K^\star$, the role of a magnetic field that makes the
coupled systems to sweep across the level-crossing point
(involving the bound $K^+$-soliton and the scattering $KN$
levels), thereby producing a resonance in analogy to resonances in
ultracold atoms driven by magnetic field~\cite{cold}.
 \eitem

I have not made any attempt to compare the bound-state formulation
based on HLS to correlated quark models~\cite{jaffe-wilczek}. A
meaningful comparison can be made only when a comprehensive
analysis is made of the spectroscopy of all pentaquqak states
predicted by the present theory. What one can say is that when a
realistic effective HLS Lagrangian is obtained from the AdS/QCD
duality, then the skyrmion description therefrom should be dual to
the quark model description. Then the debate as to which picture
is correct will be moot. This issue will be sharpened if and when
the $\Theta^+$ state is confirmed experimentally.
\subsection*{Acknowledgments}
I am grateful to Masayasu Harada and Kichi Yamawaki, the
organizers of this DSB05 meeting, for inviting me to the meeting
and for numerous discussions on the topic treated in this talk.
Part of the work discussed here was done when I was spending a
year 2004 at the Physics Department of Hanyang University. I would
particularly like to thank Professor Chong Yang Kim, the President
of Hanyang University, for the generous support that enabled me to
initiate activities in my area of research in Korea.

\end{document}